\documentstyle[preprint,revtex]{aps}

\begin{document}

\preprint{LSU preprint LSU-HE-138-1993}

\draft

\begin{title}
Structure of Abrikosov Vortices in SU(2) Lattice Gauge Theory
\end{title}

\author{V. Singh, D.A. Browne and R.W. Haymaker}

\begin{instit}
Department of Physics and Astronomy, \\
Louisiana State University, Baton Rouge, Louisiana 70803 USA
\end{instit}

\receipt{}

\begin{abstract}

We calculate the electric flux and magnetic monopole current
distribution in the presence of a static quark-antiquark pair for SU(2)
lattice gauge theory in the maximal Abelian gauge.  The current
distribution confines the flux in a dual Abrikosov vortex whose core
size is comparable to the flux penetration depth. The observed
structure is described by a dual Ginzburg-Landau model.

\end{abstract}


\narrowtext

\section{Introduction}

One explanation for the absence of free quarks is that the QCD vacuum
naturally expels color-electric flux in a manner similar to the
expulsion of magnetic flux in a superconductor.  In this
scenario~\cite{DSM,Mandel} the vacuum must contain objects that respond
to a color-electric field by generating screening currents to confine
the electric flux in a narrow tube similar to the Abrikosov flux tube
produced in a superconductor~\cite{Tink}.  While this can be achieved in
Abelian models~\cite{DSM} by adding additional Higgs fields, in the dual
superconductor model~\cite{Mandel} dynamically generated topological
excitations (magnetic monopoles) generate the currents.  In 4D U(1)
lattice gauge theory there is considerable indirect evidence from
``bulk properties'' of the vacuum such as the monopole
density~\cite{dt}, monopole susceptibility~\cite{cardy}, and static
quark potential~\cite{ws} that Dirac magnetic monopoles are associated
with the phenomenon of confinement.

In a recent paper~\cite{shb} we reported the first direct evidence that
the monopoles in U(1) lattice gauge theory actually react to the
electric flux from a static quark-antiquark pair by producing a
solenoidal current distribution to confine the flux to a narrow tube.
In the confined phase the curl of the monopole current and the local
electric field obey a dual version of the London equations~\cite{Tink}
for a superconductor.  We also found evidence for quantization of the
electric fluxoid~\cite{Tink}.  The flux tube that formed between a
static $q\bar{q}$ pair was structureless with no ``normal core''
visible on the scale of a lattice spacing.  In the deconfined phase
none of this behavior was observed.

While the origin of confinement in U(1) lattice gauge theory is fairly
clear, progress in understanding confinement in non-Abelian SU(N)
theories has been slow.  One promising approach~\cite{tHooft1} is to
fix the non-Abelian degrees of freedom in the maximal Abelian
gauge~\cite{klsw,sy}, leaving a residual U(1)$^{N-1}$ gauge freedom,
with ($N$-1) species of U(1) Dirac monopoles.  The monopoles have been
observed to be~\cite{klsw,sy,bsw,ss,ipp} abundant in the confined phase
and dilute in the (finite temperature) unconfined phase.  We present
here the first {\em direct} evidence that a dual Abrikosov vortex also
forms in SU(2) lattice gauge theory with static quarks.  We investigate
the structure of the Abrikosov vortex and find that the flux tube in
the present case has a normal core of size comparable to the flux
penetration depth.  Our results are consistent with a dual version of
the Ginzburg-Landau model of superconductivity~\cite{Tink}.  A
preliminary report of our work was presented at the LATTICE 92
conference~\cite{lattice92}.

\section{Simulations}

Our simulations were performed on a $13^3\times14$ lattice with
skew-periodic boundary conditions.  Each link from point $s$ in the
$\mu$ direction carried an SU(2) element $U(s,\mu)$ and the plaquette
operators $P_{\mu\nu}(s)$ were formed in the usual fashion as a
directed product of link variables.  We used a standard SU(2) Wilson
action
\begin{equation}
 S = \beta\sum_{s,\mu>\nu}\Bigl(1 - {1\over2}{\rm Re~Tr~} P_{\mu\nu}(s)\Bigr)
\,.
\label{Action}
\end{equation}
We generated configurations distributed according to $\exp(-S)$ using a
combination of Monte Carlo~\cite{mrrtt} and
overrelaxation~\cite{overrlx}.  Simulations were performed for
$\beta$=2.4 and 2.5 with an initial 1000 sweeps to thermalize followed
by 50 sweeps between measurements.

We converted our configurations to the maximal Abelian
gauge~\cite{tHooft1} by finding the gauge transformation that
maximized the quantity~\cite{klsw}
\begin{equation}
R = \sum_{s,\mu} {\rm~Tr~}
\Bigl[\sigma_3 U(s,\mu)\sigma_3U^{\dagger}(s,\mu)\Bigr]\,,
\label{Rgauge}
\end{equation}
which is equivalent to diagonalizing
\begin{equation}
X(s) = \sum_{\mu}\Bigl[ U(s,\mu)\sigma_3U^{\dagger}(s,\mu)
 + U(s-\mu,\mu)\sigma_3U^{\dagger}(s-\mu,\mu)\Bigr]
\label{Xgauge}
\end{equation}
at each site $s$. To measure the gauge fixing in the simulations we
used the lattice sum of the magnitude of the offdiagonal component of
$X(s)$, $|Z|^2 = \sum_s|X(s)_{12}|^2$.  Typically we needed about 650
gauge fixing sweeps to attain $|Z|^2 \approx 10^{-5}$/site for
$\beta=2.4$.  Three gauge fixing methods were used: (1) generating
and accepting random local changes only if $R$ increased, (2)
locally maximizing $R$ exactly at alternate sites, (3) applying
overrelaxation using the square of the gauge transformation of method
(2) to sample configurations better.

\section{Measurements}

After gauge fixing, the Abelian U(1) link variable is given~\cite{klsw}
by the phase of the diagonal component of the SU(2) link variable,
$u(s,\mu) = U(s,\mu)_{11}/|U(s,\mu)_{11}|$.
We construct the Abelian plaquette variables $p_{\mu\nu}(s)$ from the
$u(s,\mu)$ in the usual fashion.  The one important difference from our
U(1) study is that we have no local U(1) action here, so we cannot
identify the magnitude of the Abelian charge $e$. In the U(1) case we
divided the plaquette by $e$ in order to get the electric field
$\vec{E}$, and we multiplied the integer valued lattice operator for
$\vec{\nabla}\times \vec{J}_M$ by $e_M = 2\pi/e$ to normalize the
magnetic current.  In the present case we determine instead the
electric ``force'' $e\vec{E}$ and the magnetic number current
$2\pi\vec{J}_M/e_M = e\vec{J}_M$ as variables here, so both quantities
are known only up to an overall common factor.  We shall continue to
refer to them as the electric field and magnetic current in keeping
with our previous analysis.

Since the electric field and monopole current operators are vectors,
they will average to zero unless they are correlated with a Wilson loop
$W$ representing the current from a $q\bar{q}$ pair; a $3\times3$ loop
in the {\em z-t\/} plane was used in the simulations. Therefore,
averages of observables $\Theta$ are computed as
$\langle\Theta\rangle\equiv \mbox{Tr}\bigl\lbrace\exp(-S)
W\Theta\bigr\rbrace / \mbox{Tr}\bigl\lbrace\exp(-S) W\bigr\rbrace$.

The electric field operator is given by $a^2{\cal E}_\mu =
\mbox{Im~}p_{\mu4}$, where $a$ is the lattice spacing.  With our
choice for the orientation of the Wilson loop only the $z$-component of
the electric field has a nonzero average.  In Fig. \ref{curl}(a) we
show the operator for the electric field ${\cal E}_z$, given by a {\em
z-t\/} plaquette, as a bold line for fixed time.

The magnetic monopoles are identified using the
DeGrand-Toussaint~\cite{dt} construction.  It is convenient to
associate the monopole current density in each three-volume with a link
on the dual lattice, making world lines that define a conserved current
density $\vec{J}_M$. In Fig.~\ref{curl}(b) we show the three-volumes as
squares since the time dimension is suppressed, and through the center
of each square is the dual link associated with $\vec{J}_M$.  In order to
isolate the solenoidal monopole currents, we construct the operator for
the line integral of $\vec{J}_M$ around a dual plaquette,
$\vec{\nabla}\times \vec{J}_M$, from the four three-volumes (squares)
shown in Fig.~\ref{curl}(b).  Notice from this construction that
$\vec{E}$ and $\vec{\nabla}\times \vec{J}_M$ take values at the same
location within the unit cell of the lattice, indicated by the bold
face line in Fig.~\ref{curl}(b).

\section{Analysis}

In our previous work~\cite{shb}, we found that the confined phase of 4D
U(1) lattice gauge theory exhibited a response to a static $q\bar{q}$
pair that could be described in terms of a dual version of the London
theory~\cite{Tink}.  In this model, one combines the dual version of
Ampere's law
\begin{equation}
-c\vec{\nabla}\times\vec{E} = \vec{J}_M\,.
\label{dualAmpere}
\end{equation}
with the dual London equation governing the generation of the persistent
currents in the monopole condensate
\begin{equation}
\vec{\nabla}\times\vec{J}_M = {c\over \lambda^2} \vec{E}
\,,
\label{Londoneqns}
\end{equation}
where $\lambda = (mc^2/n_s e^2)^{1/2}$ is the London penetration depth,
with $e$, $m$ and $n_s$ the charge, mass and number density of the
monopole condensate.  The dual London theory also predicts, based on
the single-valuedness of the condensate order parameter, the fluxoid
quantization relation
\begin{equation}
\int\,\vec{E}\cdot d\vec{S} -
 {\lambda^2\over c}\oint\,\vec{J}_M\cdot d\vec{\ell} = n\Phi_e\,,
\label{fluxoid}
\end{equation}
where $n$ is an integer and $\Phi_e = e/\sqrt{\hbar c}$ is the quantum
of electric flux.

While this model explained our U(1) results, our results for SU(2) do
not obey Eqn.~(\ref{Londoneqns}).  Figure \ref{data24} shows the result
of 480 measurements at $\beta$=2.4 of $E_z$ and $(\vec{\nabla}\times
\vec{J}_M)_z$ measured midway between the $q\bar{q}$ pair.  Since
-$(\vec{\nabla}\times\vec{J}_M)_z$ has a positive value at one point
off axis, there is clearly {\it no} linear combination of the off axis
data for $E_z$ and $(\vec{\nabla}\times\vec{J}_M)_z$ that will satisfy
Eqn.~(\ref{Londoneqns}).

Therefore, we have adopted a dual form of Ginzburg-Landau (G-L)
theory\cite{Tink}, which generalizes the London theory to allow the
magnitude of the condensate density to vary in space.  As before, the
superconducting order parameter is a complex function $\psi(\vec{x})$,
where $|\psi(\vec{x})|^2$ is the condensate density.  We define
$\psi(\vec{x}) = \sqrt{n_s}f(\vec{x})\exp(i\alpha(\vec{x}))$, where
$n_s$ is the London (bulk) condensate density, and $f$ and $\alpha$ are
real functions describing the spatial variation of the condensate; the
London model presumes $f\equiv1$.  The characteristic scale over which
the condensate density varies is $\xi$, the G-L coherence length.

The dual London Eqn.~(\ref{Londoneqns}) is replaced with a dual version
of the G-L equations~\cite{Tink}
\begin{eqnarray}
\vec{E} & = & \vec{\nabla}\times\vec{A}_{E}
\nonumber \\
\vec{J}_{M} & = & {c\over\lambda^2}
f^2\Bigl(\vec{A}_E - {\Phi_e\over2\pi}\vec{\nabla}\alpha\Bigr)
\label{dualGL}
\end{eqnarray}
and the condensate density obeys
\begin{equation}
 0 = -\xi^2\nabla^2f +
\xi^2\Bigl(\vec{\nabla}\alpha-{2\pi\over\Phi_e}\vec{A}_E\Bigr)^2 f
-f + f^3\,,
\label{dualampl}
\end{equation}
while the fluxoid quantization relation of Eqn.\ (\ref{fluxoid}) is generalized
to
\begin{equation}
\int\,\vec{E}\cdot d\vec{S} -
{\lambda^2\over c}\oint\,{1\over f^2} \vec{J}_{M} \cdot d\vec{\ell}
 =  n\Phi_e\nonumber\\
\label{dualfluxoid}
\end{equation}

The dual Abrikosov vortex for a flux tube along the $z$ axis is a
solution to Eqns.~(\ref{dualAmpere}), (\ref{dualGL}) and
(\ref{dualampl}) where the GL order parameter in polar coordinates
varies as $\psi=f(r)\,e^{i\theta}$ with $f(r)$ given
approximately~\cite{Tink} by
\begin{equation}
f(r) = \tanh(0.9r/\xi).
\label{vortexsoln}
\end{equation}

To fit our data, we chose an analytic function for the azimuthal
component of the monopole current of the form
\begin{equation}
{J}_M(r) =
{a_1\over r} \left[(1+a_2r) e^{-a_2r} - (1+a_3r) e^{-a_3r}\right]
\label{fitcurl}
\end{equation}
with $a_1$, $a_2$, and $a_3$ as fitting parameters.  The form of
Eqn.~(\ref{fitcurl}) was simply chosen to ensure that $J_M(r)$ vanished
linearly at $r=0$ and faster than $1/r$ as $r\to\infty$.  The curl of
the monopole current is then found from
$(\vec{\nabla}\times\vec{J}_M(r))_z=(1/r)d(rJ_{M})/dr$ and the
parameters where fixed by fitting the data for the curl of the monopole
current.  The electric field data was then fit using
Eqns.~(\ref{vortexsoln}) and (\ref{fitcurl}) in Eqn.~(\ref{dualGL}).
We excluded the point at $r=0$ in the fit to $E_z(r)$ since we are
using a continuum approximation for the electric field to fit lattice
data for the electric flux through a plaquette, and at the origin our
continuum approximation to $E_z(r)$ diverges while the electric flux
though the plaquette is still finite.

We evaluate the fluxoid $\Phi_e$ by two distinct methods.  The first
identifies $\Phi_e$ with $\Phi_{tot}$, the net flux through the entire
lattice.  The second method comes from evaluating the fluxoid relation
on the central plaquette $(r=0)$.  If we interpret the data point
$E_z(0)$ as actually representing the flux through a circle of unit
area (radius $1/\sqrt{\pi}$), Eqn.~(\ref{dualfluxoid}) gives
\begin{equation}
\Phi_{e} \approx
\left[E_z(0) + {\lambda^2\over c} {2\pi r J_{M}(r)\over f^2(r)}
\right]_{r=1/\sqrt{\pi}}
\label{phi0}
\end{equation}

The curves in Figs.~\ref{data24}(a) and \ref{data24}(b) represent our
fit to the data for $\beta$=2.4.  The resulting current distribution as
a function of the distance from the $q\bar{q}$ axis is shown in Fig.
\ref{beta24j}.  The fitted values we obtained are $\lambda/a =
1.05\pm0.12$ and $\xi/a = 1.35\pm0.11$.  This latter result agrees
with the naive estimate that $\xi$ is the distance where
$\vec{\nabla}\times\vec{J}_M$ changes sign, {\em i.e.,\/} $\approx
1.5a$.  The fluxoid found from Eqn.~(\ref{phi0}) was $0.16\pm0.04$,
while that found from the total flux was $\Phi_{tot}=0.176\pm0.003$.
The errors quoted are the statistical variance in the fitting
parameters when the 480 measurements were divided into 4 sets of 120
points and fitting each set independently.
For $\beta=2.5$ we accumulated
208 measurements, which were grouped into 4
independent sets of 52 measurements for analysis.  We found a
penetration depth $\lambda/a=1.59\pm0.31$, a coherence length
$\xi/a=1.16\pm0.08$, and two flux quantum estimates as
$\Phi_e=0.18\pm0.03$ and $\Phi_{tot}=0.253\pm0.005$.

\section{Conclusions}

{}From this study we see that the dual superconductivity of the vacuum
in SU(2) has a richer spatial structure than that seen in U(1), with a
clearly apparent normal core region in the flux tube.  The value of
$\xi$ measures in effect the thickness of the boundary between the
external, confining vacuum and the deconfined interior of a hadron.  It
will be interesting to investigate the structure of the flux tube in
SU(3) lattice gauge theory.

The values we find for $\xi$ and $\lambda$ explain results obtained by
Ivanenko {\em et al.}~\cite{ipp} on the existence of monopoles in SU(2)
lattice gauge theory.  They found the string tension between a
$q\bar{q}$ pair correlated not with monopoles found by the
deGrand-Toussaint construction applied to elementary three-volumes, but
rather to ``extended'' monopoles defined over several three-volumes.
Since the correlation length $\xi$ measures the minimum length scale
over which a well-defined monopole condensate exists, $\xi\approx a$
means the global monopole condensate is best measured over a larger
region than an elementary three-volume.

The gross spatial structure of the flux tube is fixed by the
dimensionless Ginzburg-Landau parameter $\kappa\equiv\lambda/\xi$.  For
a type II superconductor ($\kappa>1/\sqrt{2}$) the flux tube is
compact, while a type I superconductor ($\kappa<1/\sqrt{2}$) has a
highly ramified flux structure~\cite{Huebener} in the form of
corrugated cylinders or highly reticulated walls.  Our result
($\kappa=1.4\pm0.2$) is close to the borderline.  We should point out
that an analysis by Maedan {\em et al.\/}~\cite{mms} of an effective
Lagrangian for confinement yields a value of $\kappa\gg1$ for SU(2).
However, since our value for $\kappa$ depends on the size of the Wilson
loop and on the scaling behavior of $\kappa(\beta)$, we cannot yet
reliably compare our results to a continuum model.  Further analysis
using larger Wilson loops will be needed to definitively establish the
magnitudes of $\lambda$ and $\xi$.  Rather, the goal of this work was
to find an operator $(\vec{\nabla}\times\vec{J}_M)$ that gave a clear
signal of the nature of the confinement and showed the structure of the
flux tube;  an analysis of the electric flux alone does not elucidate
this structure~\cite{Cea}.

\section{Acknowledgments}

We thank A. Kronfeld, M. Polikarpov, G. Schierholz, T. Suzuki, R.
Wensley, J. Wosiek and K. Yee for many fruitful discussions.  R.W.H.
and V.S. are supported in part by the U. S. Department of Energy under
grant DE-FG05-91ER40617 and D.A.B. is supported in part by the National
Science Foundation under grant No.\ NSF-DMR-9020310.


\figure{Operators for (a) electric field and (b) $\vec{\nabla}\times
\vec{J}_M$ on a fixed time slice.\label{curl}}

\figure{Profile of (a) $E_z$ and (b) $\vec{\nabla}\times\vec{J}_M$ as
a function of distance from the $q\bar{q}$ axis for $\beta=2.4$,
together with the best fit.\label{data24}}

\figure{Profile of the current distribution $\vec{J}_M$ derived from
the fit to the data of Fig.~\protect{\ref{data24}}.\label{beta24j}}

\end{document}